\documentclass[twocolumn,preprintnumbers,amsmath,superscriptaddress,amssymb,aps]{revtex4-1}

\usepackage{graphicx}
\usepackage{dcolumn}
\usepackage{bm}
\usepackage{amsmath}
\usepackage{amssymb}
\usepackage{graphicx}
\usepackage{indentfirst}
\usepackage{booktabs}
\usepackage{multirow}
\usepackage{colortbl}
\usepackage{float}

\selectfont

\begin{document}
\title{Type III Valley Polarization and Anomalous Valley Hall Effect in Two-Dimensional Non-Janus and Janus Altermagnet Fe$_2$WS$_2$Se$_2$}
\author{Yanchao She}
\address{Department of Physics and Electronic Engineering, Tongren University, Tongren, Guizhou, 554300, People's Republic of China}
\address{School of Physics and Mechatronics Engineering, Jishou University, Jishou, Hunan, 416000, People's Republic of China}
\author{Yiding Wang}
\address{Department of Physics and Electronic Engineering, Tongren University, Tongren, Guizhou, 554300, People's Republic of China}
\address{School of Physics and Mechatronics Engineering, Jishou University, Jishou, Hunan, 416000, People's Republic of China}
\address{State Key Laboratory for Mechanical Behavior of Materials, School of Materials Science and Engineering, Xi'an Jiaotong University, Xi'an, Shaanxi, 710049, People's Republic of China}
\author{Hanbo Sun}
\address{State Key Laboratory for Mechanical Behavior of Materials, School of Materials Science and Engineering, Xi'an Jiaotong University, Xi'an, Shaanxi, 710049, People's Republic of China}
\author{Chao Wu}
\address{State Key Laboratory for Mechanical Behavior of Materials, School of Materials Science and Engineering, Xi'an Jiaotong University, Xi'an, Shaanxi, 710049, People's Republic of China}
\author{Weixi Zhang}
\email{zhangwwxx@sina.com}
\address{Department of Physics and Electronic Engineering, Tongren University, Tongren, Guizhou, 554300, People's Republic of China}
\author{Ping Li}
\email{pli@xjtu.edu.cn}
\address{State Key Laboratory for Mechanical Behavior of Materials, School of Materials Science and Engineering, Xi'an Jiaotong University, Xi'an, Shaanxi, 710049, People's Republic of China}
\address{State Key Laboratory for Surface Physics and Department of Physics, Fudan University, Shanghai, 200433, People's Republic of China}
\address{State Key Laboratory of Silicon and Advanced Semiconductor Materials, Zhejiang University, Hangzhou, Zhejiang, 310027, People's Republic of China}

\date{\today}

\begin{abstract}
Exploiting the valley degree of freedom introduces a novel paradigm for advancing quantum information technology. Currently, the investigation on spontaneous valley polarization mainly focuses on two major types of systems. One type magnetic systems by breaking the time-reversal symmetry, the other is ferroelectric materials through breaking the inversion symmetry. Might there be additional scenarios? Here, we propose to realize spontaneous valley polarization by breaking the mirror symmetry in the altermagnets, named type III valley polarization. Through symmetry analysis and first-principles calculations, we confirm that this mechanism is feasible in Non-Janus Fe$_2$WS$_2$Se$_2$. Monolayer Non-Janus and Janus Fe$_2$WS$_2$Se$_2$ are stable N$\acute{e}$el-type antiferromagnetic state with the direct band gap semiconductor. More interestingly, their magnetic anisotropy energy exhibits the rare biaxial anisotropy and a four-leaf clover shape in the xy plane, while the xz and yz planes show the common uniaxial anisotropy. This originated from the fourth-order single ion interactions. More importantly, the valley splitting is spontaneously generated in the Non-Janus Fe$_2$WS$_2$Se$_2$ due to the M$_{xy}$ symmetry breaking, without requiring the SOC effect. Both the Non-Janus and Janus Fe$_2$WS$_2$Se$_2$ exhibit diverse valley polarization and anomalous valley Hall effect properties. In addition, the magnitude and direction of valley polarization can be effectively tuned by the biaxial strain and magnetic field. Our findings not only expand the realization system of spontaneous valley polarization, but also provide a theoretical basis for the high-density storage of valley degrees of freedom.
\end{abstract}

\maketitle
Beyond charge and spin, the valley index has rapidly gained attention as a unique electronic degree of freedom, Due to its unparalleled advantages in realizing future devices that are ultra-fast, high-capacity, power-efficient, and nonvolatile \cite{1,2,3,4}. The valley typically correspond to local energy minima or maxima points in a material's valence or conduction bands. In general, the valleys exhibit degeneracy due to symmetry connections between specific k-points, such as inversion symmetry ($\emph{P}$) and time-reversal symmetry ($\emph{T}$). To utilize the valley degrees of freedom for storage, it is necessary to break the valley degeneracy and realize valley polarization. At present, the spontaneous valley polarizations can be classified into two major categories. One category is to break the $\emph{T}$ symmetry with magnetism \cite{5,6,7,8,9}, which is named type I valley polarization. The other category is to break the $\emph{P}$ symmetry with ferroelectricity \cite{10,11}, which is type II valley polarization. Interestingly, the switching of ferroelectric polarization direction can effectively tune the valley polarization direction.

Altermagnets (AM), a newly discovered class of collinear magnets that exhibit spin splitting without net magnetization, have attracted significant attention due to their unique physical properties \cite{12,13,14,15,16,17,18}. Despite appearing as standard collinear N$\acute{e}$el-type antiferromagnet (AFM), their lattice structure actually leads to $\emph{T}$ symmetry breaking \cite{19}. Therefore, they exhibited many phenomena that distinguish from conventional AFM, such as, the ferroelectric/antiferroelectric AM switched spin splitting \cite{20,21}, the spin-polarized current generation arising from spin splitting \cite{22,23,24}, the tunneling- and giant-type magnetoresistance phenomena \cite{25,26}, the remarkable anomalous Hall effect matching ferromagnetic (FM) material \cite{27,28}, the theory-predicted spin-splitter torque with experimental validation \cite{29,30,31}, and abundant transport properties, including the spin Seebeck effect \cite{32}, crystal Nernst effect \cite{32,33}, crystal valley Hall effect \cite{34}, crystal thermal Hall effect \cite{35}, nonlinear transport \cite{36}. In addition, the valley physics in the AM has also attracted widespread attention. The valley degeneracy is protected by the mirror symmetry ($\emph{M}$) \cite{37,38}. The realization of the valley polarization requires a uniaxial strain to break the $\emph{M}$ symmetry, which is named the piezovalley effect \cite{16,17,37,39}. Whether achieve the intrinsic valley polarization without an external field in the AM?

\begin{figure*}[htb]
\begin{center}
\includegraphics[angle=0,width=1.0\linewidth]{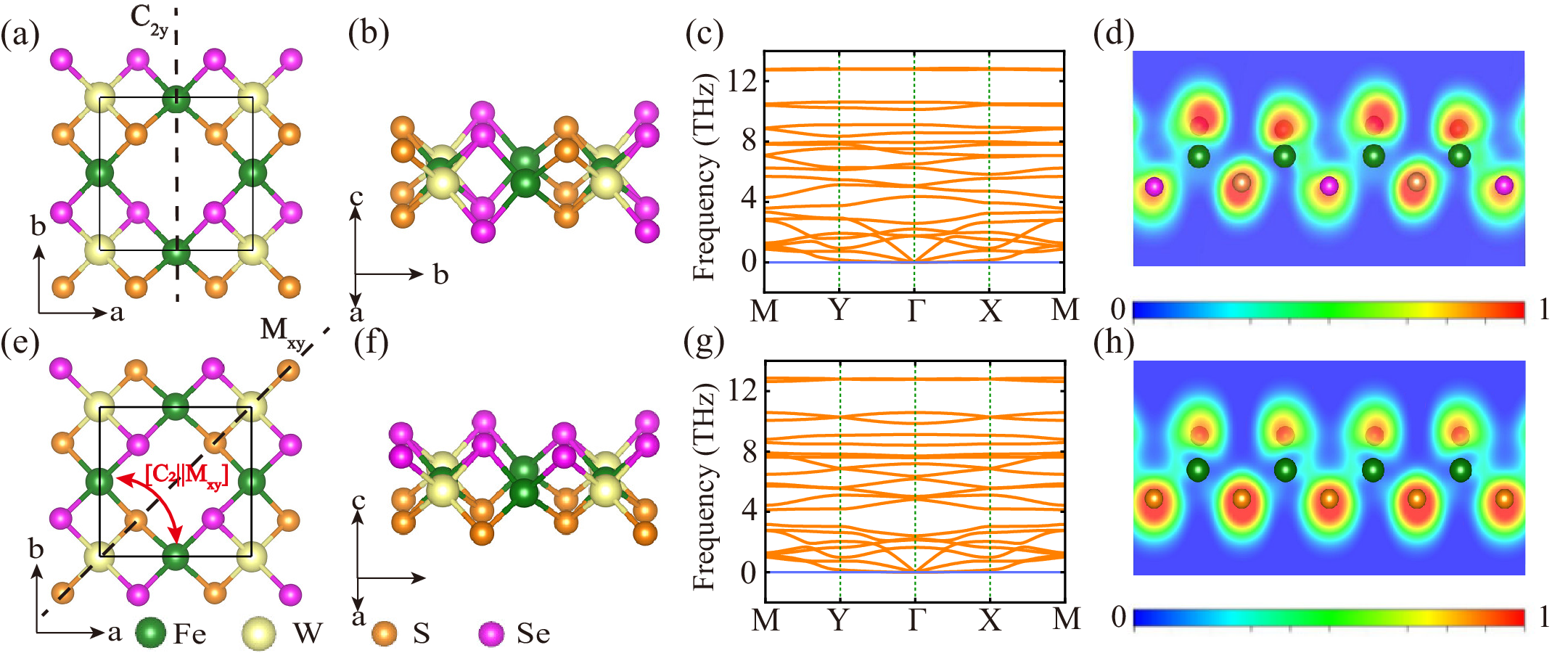}
\caption{
The top and side views of monolayer (a, b) Non-Janus and (e, f) Janus Fe$_2$WS$_2$Se$_2$ crystal structures. The green, yellow, orange, and magenta balls denote Fe, W, S, and Se atoms, respectively. (c, g) The phonon spectra of monolayer Non-Janus and Janus Fe$_2$WS$_2$Se$_2$, respectively. (d, h) Electron localization function of monolayer Non-Janus and Janus Fe$_2$WS$_2$Se$_2$, respectively.
}
\end{center}
\end{figure*}

In this work, based on symmetry analysis, we design the Non-Janus structure to break the $\emph{M}$ symmetry in two-dimensional (2D) AMs, realizing the intrinsic valley polarization. Since it breaks the $\emph{M}$ symmetry, we name it the type III valley polarization. Through density functional theory (DFT) calculations, the feasibility of this mechanism is confirmed in the Fe$_2$WS$_2$Se$_2$ system. We have predicted two structures: monolayer Non-Janus and Janus Fe$_2$WS$_2$Se$_2$. Our results show that both materials are direct band gap semiconductors of the N$\acute{e}$el type AFM state. More interestingly, the magnetic anisotropy energy (MAE) presents a four-leaf clover shape in the xy plane, while the xz and yz planes exhibit an eight-shaped pattern. Their realization of valley polarization all originates from the symmetry breaking of the M$_{xy}$ crystal, but the methods are completely different. The monolayer Non-Janus Fe$_2$WS$_2$Se$_2$ is achieved through the structure, while monolayer Janus Fe$_2$WS$_2$Se$_2$ is realized by in-plane magnetization. Moreover, the biaxial strain and magnetization direction can effectively tune the magnitude and direction of valley polarization, and realize the anomalous valley Hall effect. Our findings have expanded the ways in which valley polarization can be achieved.

\subsection*{Results}	
\subsection{Structure, symmetry and stability}
Monolayer Non-Janus and Janus Fe$_2$WS$_2$Se$_2$ has a tetragonal lattice with the space group $\emph{P2}$, $\emph{Cmm2}$ and point group C$_2$, C$_{2v}$, respectively. As shown in Fig. 1(a, b, e, f), they both consist of three atomic layers, where the Fe-W atomic layer is sandwiched by S, Se, or S-Se atomic layers. Since the S and Se atoms separate two Fe atoms, there's practically no overlapping of electron clouds between neighboring Fe atoms, thereby preventing direct exchange interaction. The lattice constants of monolayer Non-Janus and Janus Fe$_2$WS$_2$Se$_2$ are 5.49 ${\rm \AA}$ and 5.51 ${\rm \AA}$, which they are almost unanimous. However, the total energy of Non-Janus Fe$_2$WS$_2$Se$_2$ is 64.74 meV lower than that of Janus Fe$_2$WS$_2$Se$_2$. Then, we evaluate the dynamical stability of monolayer Non-Janus and Janus Fe$_2$WS$_2$Se$_2$ by calculating the phonon spectra. As shown in Fig. 1(c, g), the absence of imaginary frequencies in Non-Janus and Janus Fe$_2$WS$_2$Se$_2$ indicates that their dynamics are stable. To compare the differences in bonding between Non-Janus and Janus structures, we calculate the electron localization function (ELF). As shown in Fig. 1(d, h), the electrons are mainly localized around the S and Se atoms, indicating typical ionic bonding for the Fe-S and Fe-Se bond. It is worth noting that there is a weak electron cloud connecting S and Se atoms through the Fe atoms. It indicates the presence of weak Fe-S and Fe-Se covalent bond characteristics. The covalent bond of Non-Janus Fe$_2$WS$_2$Se$_2$ is slightly stronger than that of Janus Fe$_2$WS$_2$Se$_2$.

\begin{figure*}[htb]
\begin{center}
\includegraphics[angle=0,width=1.0\linewidth]{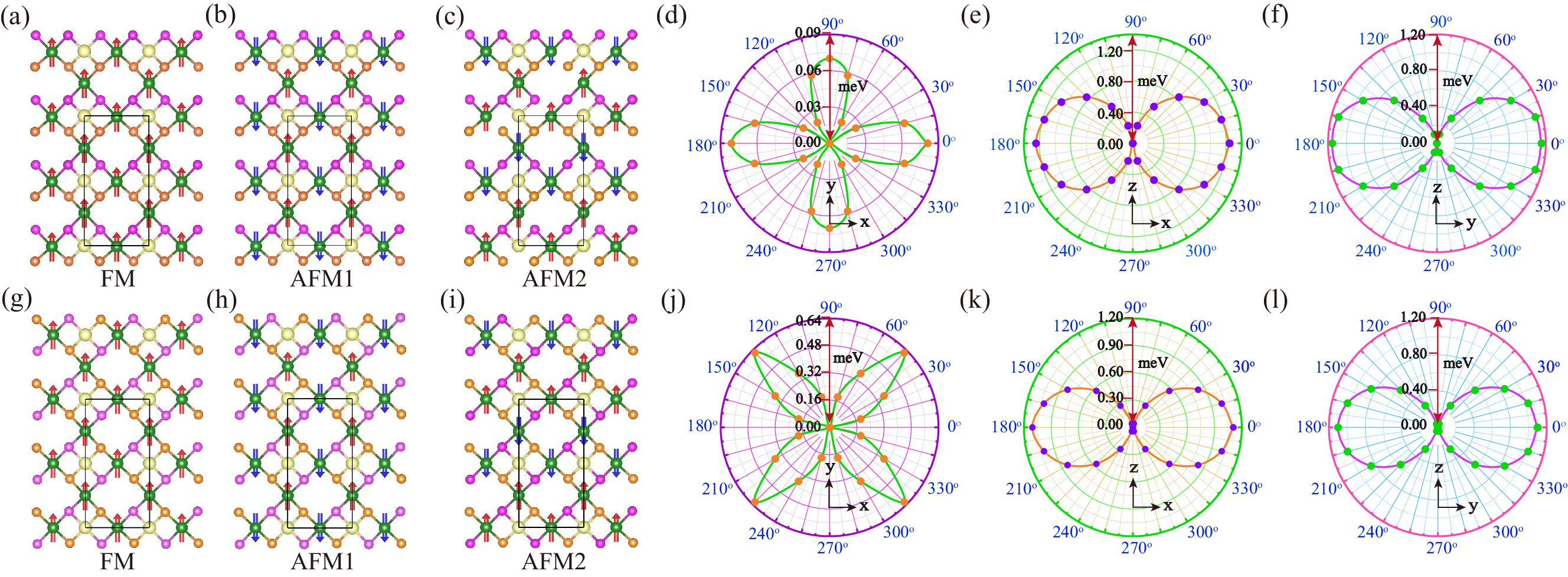}
\caption{
(a-c) and (g-i) represent three magnetic configurations of monolayer Non-Janus and Janus Fe$_2$WS$_2$Se$_2$, respectively. Angular dependence of the MAE of (d, e, f) monolayer Non-Janus and (j, k, l) Janus Fe$_2$WS$_2$Se$_2$ with the direction of magnetization lying on the xy, xz, and yz plane, respectively.
}
\end{center}
\end{figure*}

The two sublattices have no symmetrical connection in Non-Janus Fe$_2$WS$_2$Se$_2$, while two sublattices of Janus Fe$_2$WS$_2$Se$_2$ are linked through the M$_{xy}$ mirror symmetry. In addition, Non-Janus and Janus Fe$_2$WS$_2$Se$_2$ also posses the C$_{2y}$ and C$_{2z}$ rotation symmetry, respectively. Since the S and Se atoms of Non-Janus Fe$_2$WS$_2$Se$_2$ are equal in the upper and lower layers, there is no potential difference in the z direction (see Fig. S1(a)). On the contrary, as shown in Fig. S1(b), Janus Fe$_2$WS$_2$Se$_2$ has an electric potential difference of 0.19 eV. Therefore, the built-in electric field is 0.07 V/${\rm \AA}$ \cite{40,41}.

\subsection{Magnetic property}
To confirm the magnetic ground states of monolayer Non-Janus and Janus Fe$_2$WS$_2$Se$_2$, as shown in Fig. 2(a-c, g-i), three possible magnetic configurations are considered in a $1\times 2\times 1$ supercell, including the FM, N$\acute{e}$el-type AFM (named AFM1), and zigzag-type AFM (named AFM2) states. The calculated results are summarized in Table SI, which indicates that AFM1 is the magnetic ground state for both Non-Janus and Janus Fe$_2$WS$_2$Se$_2$. For the crystal structure, they both lack the $\emph{P}$ symmetry. Simultaneously, the Janus Fe$_2$WS$_2$Se$_2$ has [C$_2$$\|$M$_{xy}$] spin symmetry, while Non-Janus has no [C$_2$$\|$M$_{xy}$] spin symmetry. It indicates that Janus Fe$_2$WS$_2$Se$_2$ possess the characteristics of $\emph{d}$-wave altermagnet.

The importance of MAE in spintronics is well recognized, being a fundamental characteristic of magnetic materials. The Mermin-Wagner theorem dictates that the MAE is indispensable for maintaining long-range magnetic order in 2D systems \cite{42}. Hence, we investigate the MAE, defined as MAE = E$_{001}$ - E$_{100}$, representing the energy difference between the magnetic orientations along the [001] and [100] crystallographic directions \cite{43,44}. The MAE of Non-Janus and Janus Fe$_2$WS$_2$Se$_2$ are 1.24 meV and 1.11 meV, respectively. This means that the easy magnetization direction are both along the z axis. In order to the further understanding of the distribution of magnetic anisotropy throughout the entire space, we calculate the MAE in the xy, xz, and yz planes. As shown in Fig. 2 (d-f, j-l), it is a typical uniaxial MAE in the xz and yz planes. It can be well described by the equation
\begin{equation}
\rm MAE= K_1 cos^2\theta + K_2 cos^4\theta,
\end{equation}
However, the MAE in the xy plane cannot be described by the equation 1. It is a biaxial MAE that is rarely seen in real material systems. Its Hamiltonian needs to be written in a fourth-order form \cite{45}
\begin{equation}
H_{\rm MAE}= \sum_i A_4(S_{ix}^2S_{iy}^2+S_{ix}^2S_{iz}^2+S_{iy}^2S_{iz}^2),
\end{equation}
where A$_4$ denotes fourth-order single ion anisotropy (SIA). As shown in Fig. 2 (d, j), the MAE exhibits four degenerate minima (forming four-leaf clover shape) while spins rotate in xy plane. The four degenerate energy minimum appear along the [110] and [1$\bar{1}$0] directions for the Non-Janus Fe$_2$WS$_2$Se$_2$, while the four degenerate energy minimum of Janus Fe$_2$WS$_2$Se$_2$ occurs along the [100] and [010] directions. The direction of four degenerate energy minimum is entirely determined by the positions of S and Se atoms. The S and Se atoms of Non-Janus Fe$_2$WS$_2$Se$_2$ are horizontally distributed, while those of Janus Fe$_2$WS$_2$Se$_2$ are arranged along the diagonal. More importantly, the biaxial MAE can be encoded into higher order logic bits for the biaxial magnetic tunnel junction (e.g., encoding [00], [01], [10], and [11] four states).

\begin{figure*}[htb]
\begin{center}
\includegraphics[angle=0,width=1.0\linewidth]{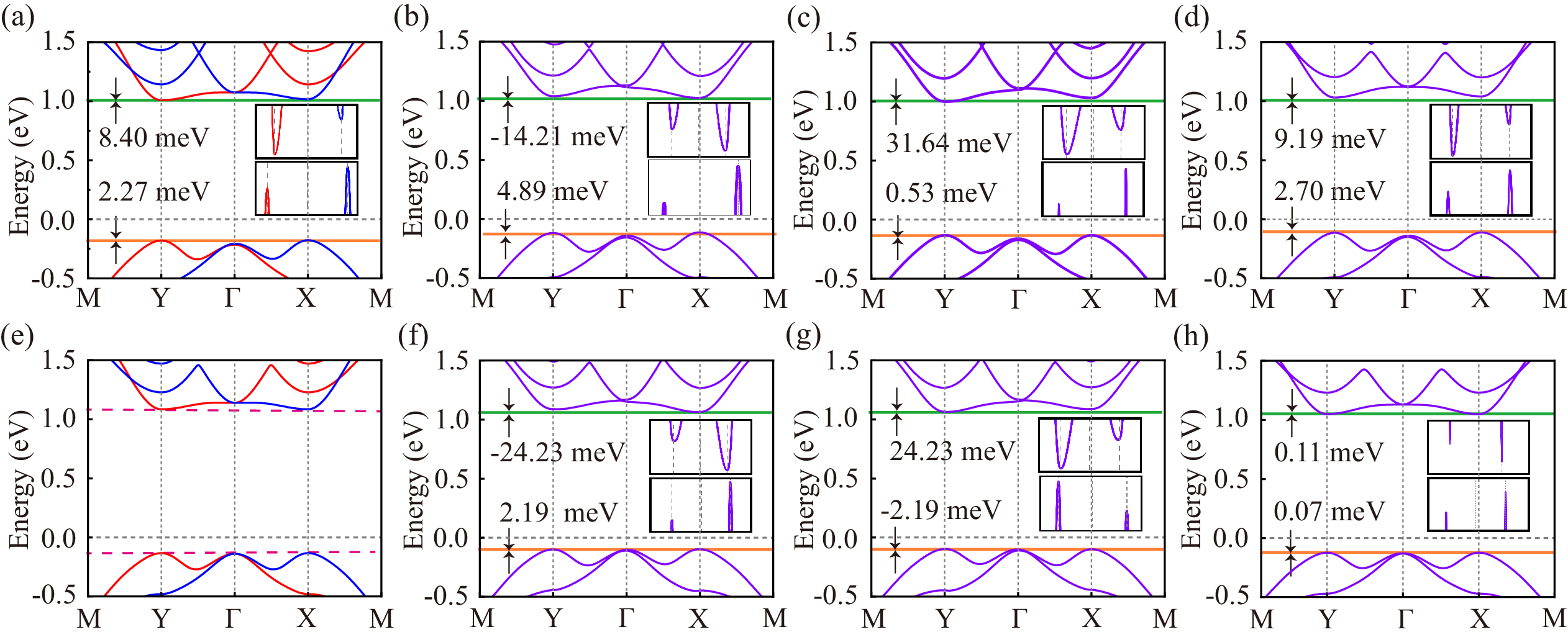}
\caption{
Band structures of monolayer (a-d) Non-Janus and (e-h) Janus Fe$_2$WS$_2$Se$_2$. (a, e) Spin-polarized band structures, (b, f) with the M//x SOC effect, (c, g) with the M//y SOC effect, and with the M//z SOC effect. The red and blue lines denote spin up and spin down bands, respectively. The valence and conduction bands valley splitting are shown by the orange and green shadings, respectively.
}
\end{center}
\end{figure*}

\subsection{Electronic properties and valley polarization}
In the following, we explore the band structures of monolayer Non-Janus and Janus Fe$_2$WS$_2$Se$_2$, and compare the differences between them due to their different symmetries. As shown in Fig. 3(a, e), when the spin-orbit coupling (SOC) is switched off, they are both direct band gap semiconductors with the valence band maximum (VBM) and conduction band minimum (CBM) at the X and Y points.
The VBM and CBM at the X point are spin down band, while the VBM and CBM at Y point are spin up band. Both exhibit the characteristic of spin splitting. Since the Janus Fe$_2$WS$_2$Se$_2$ has [C$_2$$\|$M$_{xy}$] symmetry, the VBM and CBM at the X and Y points are degenerate in terms of energy. However, when Janus Fe$_2$WS$_2$Se$_2$ transforms to Non-Janus Fe$_2$WS$_2$Se$_2$, the symmetry of [C$_2$$\|$M$_{xy}$] is broken. As shown in Fig. 3(a), the energy degeneracy of VBM and CBM at the X and Y points disappears, leading to a 2.27 meV and 8.40 meV valley splitting in the VBM and CBM, respectively. Here, the valley splitting is defined as E$_{\rm V}$(E$_{\rm C}$) =  E$_{\rm XC(V)}$ - E$_{\rm YC(V)}$, where the E$_{\rm XC(V)}$ and E$ _{\rm YC(V)}$ denote the VBM(CBM) energy at the X point and the VBM(CBM) energy at the Y point. The valley splitting of Non-Janus Fe$_2$WS$_2$Se$_2$ is entirely caused by structural symmetry, unlike the previously reported valley polarization that is triggered by the broken $\emph{P}$ or $\emph{T}$ symmetry \cite{5,6,7,8,9,10,11}. Therefore, the valley splitting caused by the broken structural symmetry that is named type III valley polarization.

\begin{figure*}[htb]
\begin{center}
\includegraphics[angle=0,width=1.0\linewidth]{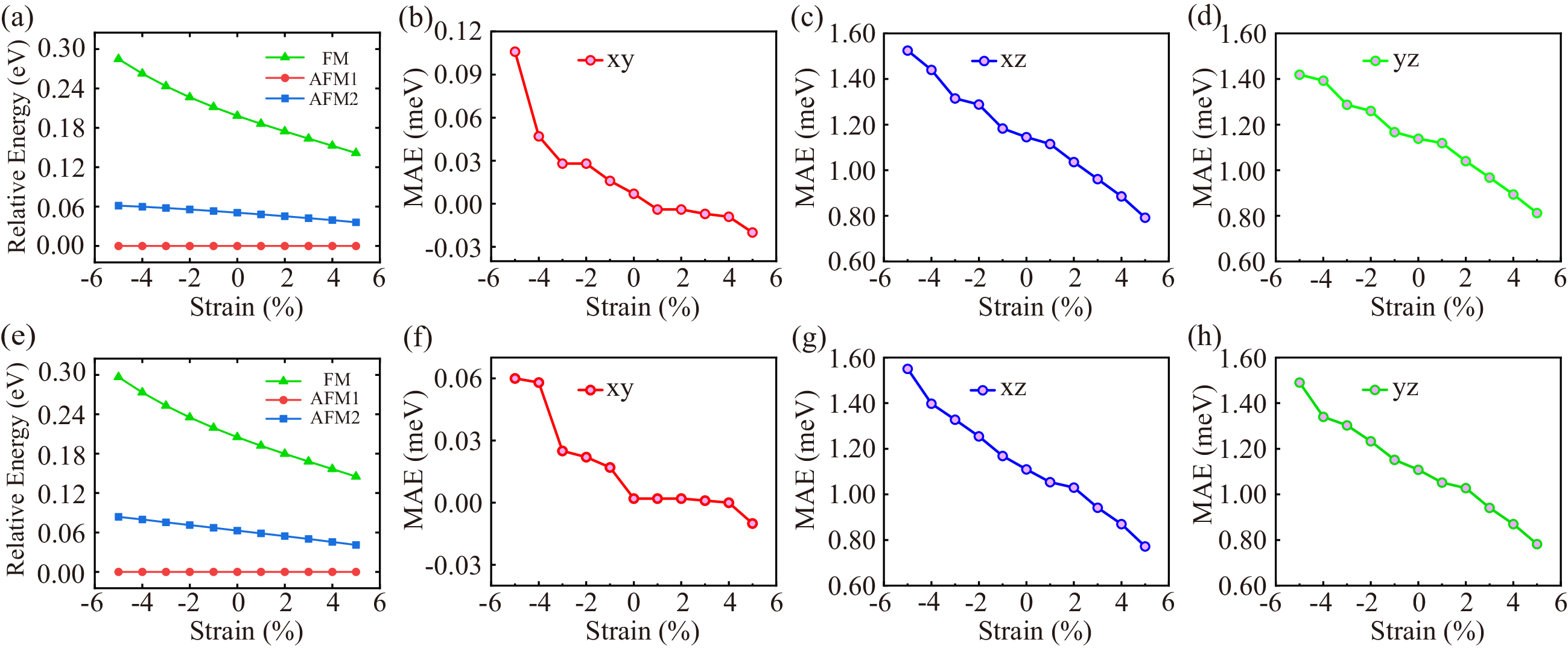}
\caption{(a, e) Calculated the total energies of different magnetic configurations for monolayer (a) Non-Janus and (e) Janus Fe$_2$WS$_2$Se$_2$ under different biaxial strain. The MAE of monolayer (b-d) Non-Janus and (f-h) Janus Fe$_2$WS$_2$Se$_2$ under different biaxial strains. (b, f), (c, g), and (d, h) exhibit the energy difference E$_x$ - E$_y$, E$_x$ - E$_z$, and E$_y$ - E$_z$, respectively. Where E$_x$, E$_y$, and E$_z$ denote the total energy of system, when the magnetization direction is along the x, y, or z axis.
}
\end{center}
\end{figure*}

\begin{figure*}[htb]
\begin{center}
\includegraphics[angle=0,width=1.0\linewidth]{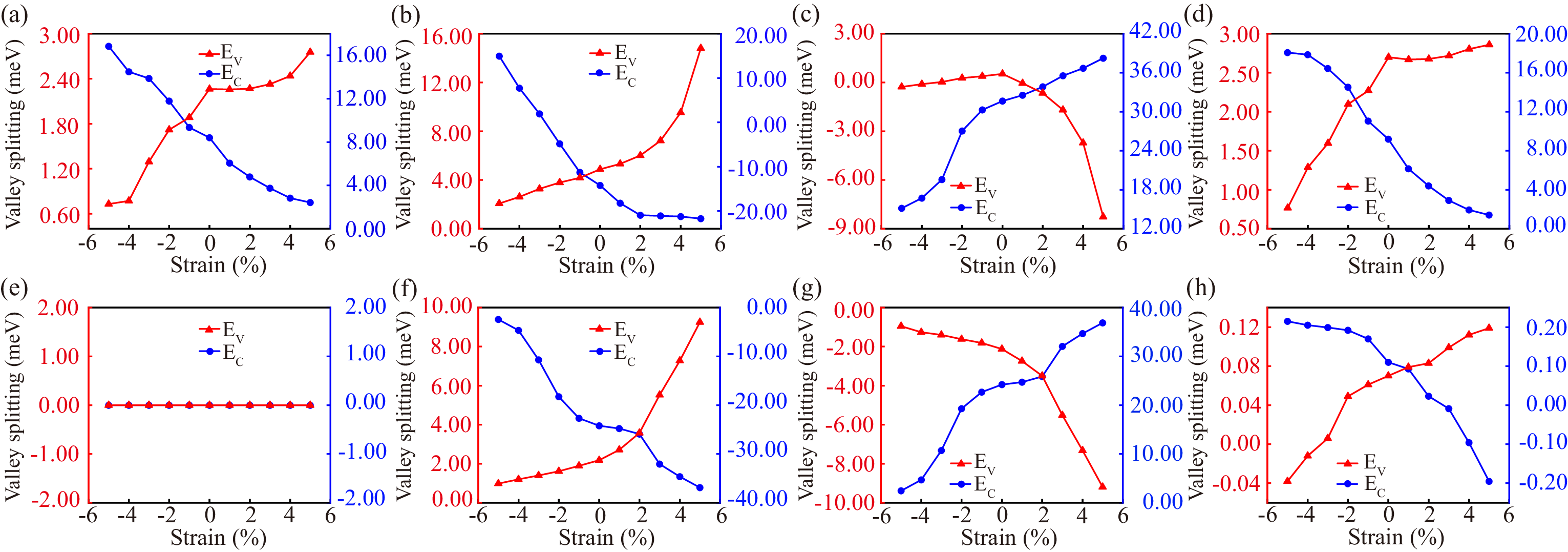}
\caption{ The valley splitting of monolayer (a-d) Non-Janus and (e-h) Janus Fe$_2$WS$_2$Se$_2$ (a, e) without the SOC effect, (b, f) with the SOC effect in M//x, (c, g) with the SOC effect in M//y, and (d, h) with the SOC effect in M//z. E$\rm_V$ and E$\rm_C$ denote the valence and conduction valley splitting, respectively.
}
\end{center}
\end{figure*}

When the SOC is included, as shown in Fig. 3(b-d, f-h), they all occur the valley splitting for the magnetization direction along +x axis, +y axis, and +z axis. Unlike the hexagonal lattice, the valley polarization often arise in the out-of-plane magnetization \cite{5,6,7,8}. For the Janus Fe$_2$WS$_2$Se$_2$, when the magnetization is along +x axis, the 2.19 meV and -24.23 meV valley splitting appear in the VBM and CBM due to the broken [C$_2$$\|$M$_{xy}$] symmetry. The significant difference of valley splitting at the VBM and CBM is caused by the compositions of bands. As shown in Fig. S2, the VBM bands are mainly composed of Fe d$_{xz}$ and d$_{yz}$ orbitals, while the CBM bands are dominated by Fe d$_{x^2-y^2}$ and d$_{z^2}$ orbitals. As is well known, the d$_{x^2-y^2}$ orbital lies on the xy plane, while the d$_{xz}$ and d$_{yz}$ orbitals have certain angles with the xy plane. Therefore, the breaking of [C$_2$$\|$M$_{xy}$] symmetry has a greater affect on the valley splitting of CBM bands. When the magnetization direction is switched from +x axis to +y axis, as shown in Fig. 3(g), the valley splitting also are reversed. However, when the magnetization direction is along the +z axis, as shown in Fig. 2(h), the valley splitting can be disregarded due to the presence of [C$_2$$\|$M$_{xy}$] symmetry. For the Non-Janus Fe$_2$WS$_2$Se$_2$, the valley splitting in all three magnetization directions is quite significant due to the broken [C$_2$$\|$M$_{xy}$] symmetry. Nevertheless, the larger valley splitting of the CBM bands compared to the VBM bands due to the composition of the orbitals (see Fig. S2). When the magnetization direction is along x axis and y axis, as shown in Fig. 3(b, c) the results of valley splitting are completely different. For the x axis case, the valley splitting of VBM and CBM has significantly increased, and the valley polarization direction of CBM has also changed. When the magnetization direction changes to the y axis, the valley splitting of VBM has decreased. It's worth noting that the valley polarization direction of CBM is further switched and increased. While the magnetization direction is along the z axis, as shown in Fig. 3(d), the magnitude of valley splitting is almost the same as that without the SOC effect. The extremely rich valley splitting phenomenon in different magnetization directions is of great significance for the practical application of valley physics.

\subsection{Strain tuned magnetic properties and valley splitting}

\begin{figure*}[htb]
\begin{center}
\includegraphics[angle=0,width=1.0\linewidth]{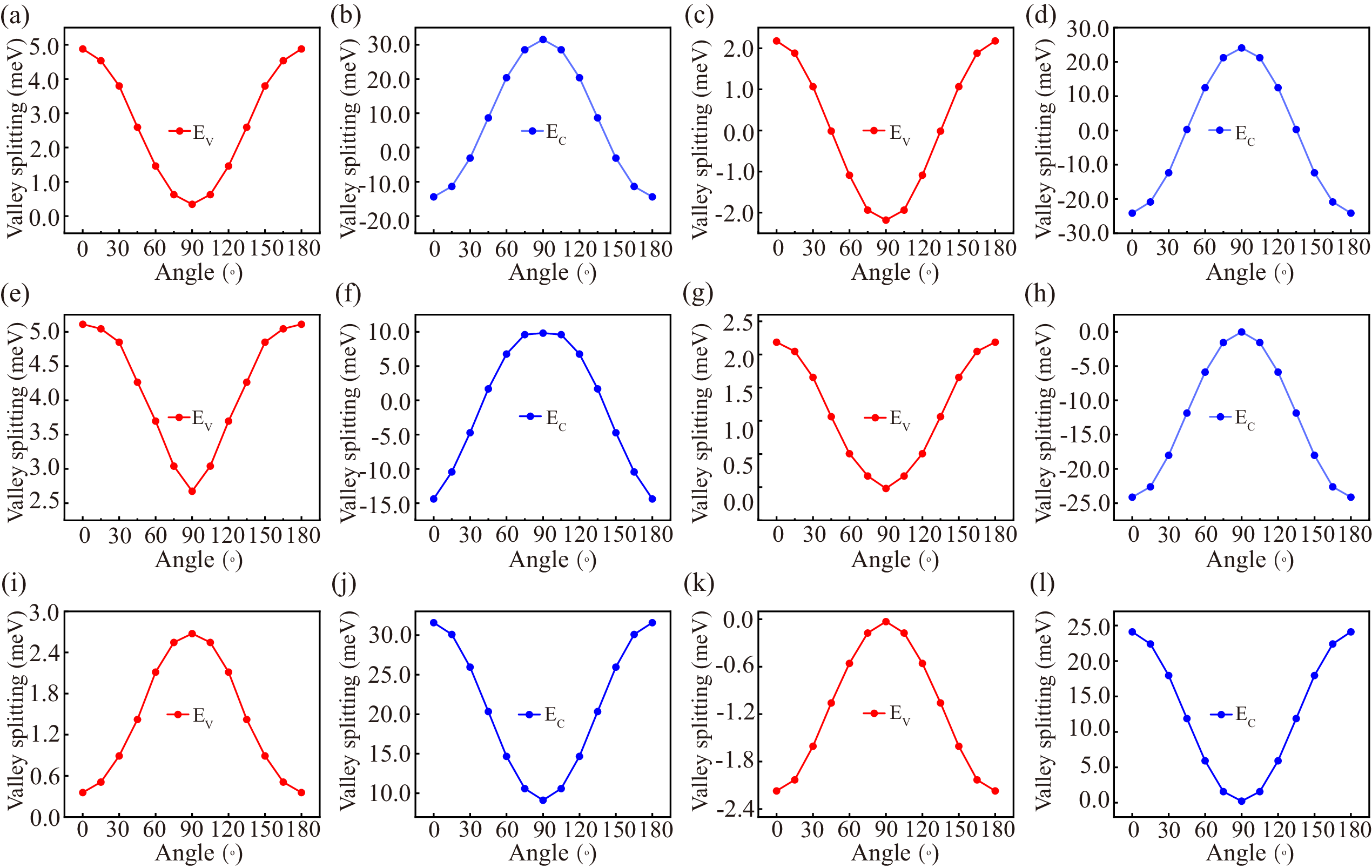}
\caption{ Magnetization direction dependence of the valley splitting of monolayer (a, b, e, f, i, j) Non-Janus and (c, d, g, h, k, l) Janus Fe$_2$WS$_2$Se$_2$ in (a-d) xy, (e-h) xz, and (i-l) yz planes. E$\rm_V$ and E$\rm_C$ denote the valence and conduction valley splitting, respectively.
}
\end{center}
\end{figure*}

Next, we investigate the magnetic properties and valley splitting of monolayer Non-Janus and Janus Fe$_2$WS$_2$Se$_2$ under the -5$\%$ $\sim$ 5$\%$ biaxial strains. The biaxial strain is defined as $\varepsilon$ = (a-a$_0$)/a$_0$$\times$100$\%$, where a$_0$ and represent the lattice constant before and after the in-plane biaxial strain, respectively. Fig. 4(a, e) exhibits the total energy of Non-Janus and Janus Fe$_2$WS$_2$Se$_2$ changes for the three magnetic configurations under the biaxial strain. The energy of AFM1 state is much lower than that of the FM and AFM2 magnetic configurations (see Table SII and SIII). It indicate that the AFM1 ground state remains extremely resistant to the biaxial strains. Moreover, as shown in Fig. 4(b-d, f-h), the strain has a relatively minor impact on MAE. Interestingly, the MAE of xy, xz, and yz planes trends are consistent for Non-Janus and Janus Fe$_2$WS$_2$Se$_2$ (see Table SIV and SV). Their MAE gradually decreases from compressive strain to tensile strain. Within the strain range of -5$\%$ $\sim$ 5$\%$, the out-of-plane easy magnetization axis was still maintained. This is crucial for 2D magnetic systems. Since the growth of 2D magnetic materials cannot be achieved without the support of a substrate. The substrate inevitably causes in-plane strain for 2D magnetic materials. The AFM1 ground state and out-of-plane magnetization under the biaxial strain, which is conducive to experimental verification that the Non-Janus and Janus Fe$_2$WS$_2$Se$_2$ possess abundant AM and valley polarization properties.

We systematically investigate the valley splitting of Non-Janus and Janus Fe$_2$WS$_2$Se$_2$ without and with the SOC effect under the biaxial strain. As shown in Fig. 5, the valley splitting can be significantly tuned under the biaxial strain. In the absence of SOC, as shown in Fig. 5(a), the VBM valley splitting of Non-Janus Fe$_2$WS$_2$Se$_2$ gradually becomes larger from compressive strain to tensile strain, while the valley splitting of CBM has decreased from 16.84 meV to 2.43 meV (see Table SVI). The band structures are shown in Fig. S4. On the contrary, the Janus Fe$_2$WS$_2$Se$_2$ will not occur valley splitting under the biaxial strain (see Fig. 5(e)). Since it requires breaking $\emph{T}$ symmetry combined the SOC effect to realize valley splitting, that belongs to the type I valley polarization. When the SOC is included, the variation trend of Non-Janus and Janus Fe$_2$WS$_2$Se$_2$ are same under the biaxial strain. When the magnetization direction is along the x axis, as shown in Fig. 5(b, f), their valence valley splitting show a parabolic increase, while the conduction valley splitting decreases almost linearly. It is worth noting that the variation range of the conduction valley splitting is as high as $\sim$40 meV. When the magnetization direction switches from x axis to y axis, as shown in Fig. 5(c, g), the trend of the valley splitting has been completely reversed. This originates from the spin-valley locking. When the magnetization direction changes to the out-of-plane, as shown in Fig. 5(d, h), the trend of valley splitting and the magnetization direction along the x axis are consistent. Here, it should be noted that the valley splitting of Janus Fe$_2$WS$_2$Se$_2$ is extremely small and can be disregarded due to the [C$_2$$\|$M$_{xy}$] symmetry. All band structures and valley splitting values are shown in Fig. S7-S10, Table SVI, and Table SVII.

\subsection{Magnetic tuned valley splitting}
Monolayer Non-Janus and Janus Fe$_2$WS$_2$Se$_2$ have spin splitting and valley polarization, which is typical of a multiferroic material. Unlike typical multiferroics, which combine ferroelectricity and magnetism, these materials exhibit intertwined altermagnetic and ferrovalley orders. The electronic properties of these multiferroic materials in response to magnetic fields are highly worthy of study. Therefore, we investigate how valley splitting depends on magnetization orientation. As shown in Fig. 6, it shows the valley polarization varies of the VBM and CBM bands with the magnetization direction. On the whole, the external magnetic field can significantly tune the valley splitting, that they present a perfect parabolic curve as the magnetization direction changes. Noted that the valley splitting trends of the valence and conduction bands are exactly opposite for the magnetization direction in the xy, xz, and yz planes. More importantly, it is different from the FM Cr$_2$COOH and AFM MnBr that we previously reported \cite{7,41}, the valley splitting and MAE are both uniaxial anisotropy. However, the MAE of Non-Janus and Janus Fe$_2$WS$_2$Se$_2$ exhibit biaxial anisotropy in the xy plane, while the valley splitting shows uniaxial anisotropy. This asynchrony is entirely determined by the band structure resulting from the spin splitting of the altermagnet. On the contrary, the valley splitting and MAE exhibit consistent uniaxial anisotropy for xz and yz planes. The valley splitting values are listed in Table SX and SXI.

\subsection{Anomalous valley Hall effect}

\begin{figure}[htb]
\begin{center}
\includegraphics[angle=0,width=1.0\linewidth]{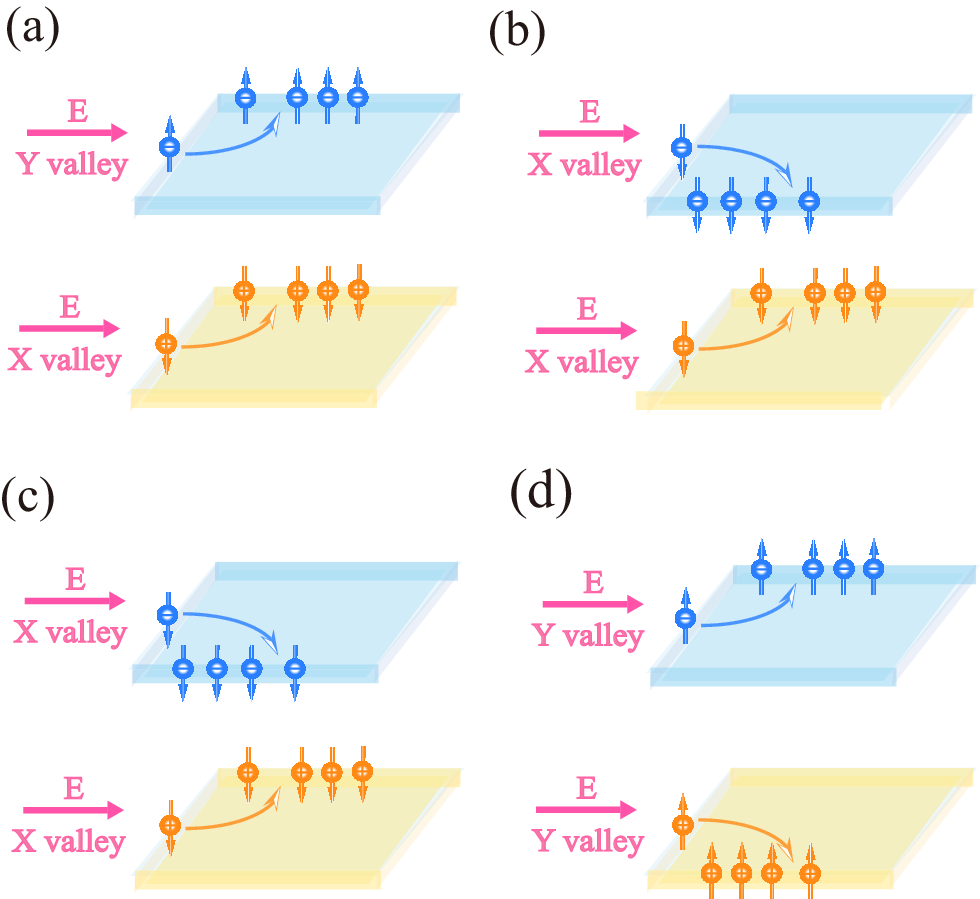}
\caption{ Schematic diagram of anomalous valley Hall effect monolayer (a) for the Non-Janus Fe$_2$WS$_2$Se$_2$ without the SOC effect and with the SOC effect in M//y and M//z, (b) with the SOC effect in M//x, (c) for the monolayer Janus Fe$_2$WS$_2$Se$_2$ with the SOC effect in M//x and M//z, and (d) with the SOC effect in M//y. The + and - represent the holes and electrons, respectively. The spin up and spin down carriers are denoted by upward and down ward arrows, respectively.
}
\end{center}
\end{figure}

From the perspective of valleytronic device applications, the understanding microscopic mechanism of the anomalous valley Hall effect (AVHE) is crucial importance. Therefore, we drew the AVHE schematic diagram in the hole and electron doping cases for the Non-Janus and Janus Fe$_2$WS$_2$Se$_2$, as shown in Fig. 7. In the absence of SOC case, when the Fermi level is tuned between the X and Y valleys in the CBM band, the spin up electrons of Y valley will be generated and accumulate on the left boundary of Non-Janus and Janus Fe$_2$WS$_2$Se$_2$ under an in-plane electrical field [see upper plane of Fig. 7(a)]. While the Fermi level is moved  between the X and Y valleys in the VBM band, as shown in Fig. 7(a) lower plane, it will become the spin down hole accumulating on the left boundary. When the SOC (M//x) is included, with the electron-doping condition, the spin down electrons from X valley will be produced and accumulate on the right boundary [see upper plane of Fig. 7(b)]. When it becomes hole-doping, as shown in Fig. 7(b) lower plane, the spin down hole will be observed on the left boundary. While the magnetization direction is along the y axis and z axis, the image of AVHE is consistent with that without the SOC effect. Moreover, the AVHE of Janus Fe$_2$WS$_2$Se$_2$ with the SOC in M//x and M//y is the same as that of Non-Janus Fe$_2$WS$_2$Se$_2$ with the SOC in M//x, as shown in Fig. 7(c). When the magnetization direction is along the y axis, all the circumstances are completely reversed compared to the magnetization direction in x axis and z axis [see Fig. 7(d)]. The abundant AVHE offers more options for the application of valleytronics.

\subsection*{Discussion}
In conclusion, we propose that the type III valley polarization realizes by breaking the $\emph{M}$ symmetry. Based on first-principles calculation, the mechanism is confirmed in Non-Janus Fe$_2$WS$_2$Se$_2$. Meanwhile, we compare the type I valley polarization Janus Fe$_2$WS$_2$Se$_2$ resulting form the broken $\emph{T}$ symmetry. Our results show that both Non-Janus and Janus Fe$_2$WS$_2$Se$_2$ are stable AFM1 ground states with the spin splitting in the absence of SOC, and exhibit the direct band gap semiconductors. More importantly, we find biaxial MAE in real materials xy plane, which is extremely rare. This originated from the fourth-order single ion interactions. While the xz and yz planes exhibit the common uniaxial MAE. The Non-Janus Fe$_2$WS$_2$Se$_2$ spontaneously produces valley splitting without the SOC effect due to the broken M$_{xy}$ symmetry. When the SOC is included, the Non-Janus and Janus Fe$_2$WS$_2$Se$_2$ demonstrated a rich variety of valley polarization and AVHE properties. Moreover, the strain and magnetic field can effectively tune the magnitude and direction of valley polarization. Our work have provided a direction for high-density storage based on the valley degree of freedom.

\subsection*{Methods}
We use, based on the density functional theory (DFT), the Vienna $Ab$ $initio$ Simulation Package (VASP) to investigate the magnetic ground state and electronic properties\cite{46,47,48}. The exchange-correlation energy is described employing the generalized gradient approximation (GGA) with the Perdew-Burke-Ernzerhof (PBE) functional\cite{49}. A plane-wave basis set with a kinetic energy cutoff of 600 eV is used. The $\Gamma$-centered k-meshes of $12\times 12\times 1$ and $20\times 20\times 1$ are employed for structural relaxation and self-consistent calculations, respectively. To eliminate artificial interactions between periodic images, a 20 $\rm \AA$ vacuum spacing is applied along the c-axis. The calculations used force and energy convergence criteria of -0.001 eV/$\rm \AA$ and 10$^{-6}$, respectively. The GGA+U method is used to account for strong electron correlations in Fe 3d orbitals\cite{50,51,52,53}, with a Coulomb parameter U$_{eff}$ = 3 eV. The dynamical stability is evaluated through phonon spectrum calculations implemented in PHONOPY code, utilizing a $3\times 3\times 1$ supercell\cite{54}.

\subsection*{Data availability}
The data that support the findings of this study are available from the corresponding author upon reasonable request.

\subsection*{Code availability}
The codes are available from the findings of this study are available from the corresponding author on reasonable request.


\subsection*{Acknowledgements}
This work is supported by the National Natural Science Foundation of China (Grants No. 12474238, and No. 12004295), the Natural Science Foundation of Guizhou Provincial Education Department of China (Grant No. ZK[2022]558, ZK[2023]467), P. Li also acknowledge supports from the China's Postdoctoral Science Foundation funded project (Grant No. 2022M722547), the Fundamental Research Funds for the Central Universities (xzy012025031), the Open Project of State Key Laboratory of Surface Physics (No. KF2024$\_$02), and the Open Project of State Key Laboratory of Silicon and Advanced Semiconductor Materials (No. SKL2024-10). Y. She acknowledge supports from the NSF of Tongren Science and Technology Bureau (Grants No. [2023]41).

\subsection*{Author contributions}
P. L. conceived the idea. Y. S. and Y. W. performed the theoretical calculation and numerical simulation. Y. S. and Y. W. prepared the figures. Y. S., Y. W., W. Z., and P. L. did data analysis. P. L. wrote the paper.

\subsection*{Competing interests}
The authors declare no competing interests.

\subsection*{Additional information}
\textbf{Supplementary information} The online version contains supplementary material available.

\end{document}